\documentclass[conference]{IEEEtran}

\usepackage{csquotes}
\usepackage{graphicx}

\usepackage{enumitem}
\usepackage[para,online,flushleft]{threeparttable}

\usepackage{subcaption}
\usepackage{booktabs}

\usepackage[hyphens]{url}
\usepackage{hyperref}

\newcommand{\specialcell}[2][c]{%
	\begin{tabular}[#1]{@{}c@{}}#2\end{tabular}}

\author{}
\date{}

\IEEEoverridecommandlockouts

\begin{document}

\title{\textbf{\sffamily Combinatorial Optimization of Work Distribution on Heterogeneous Systems}}

\author{\IEEEauthorblockN{Suejb Memeti and
		Sabri Pllana\\}
	\IEEEauthorblockA{Department of Computer Science,
		Linnaeus University\\
		351 95 V\"{a}xj\"{o}, Sweden\\
		\{suejb.memeti, sabri.pllana\}@lnu.se}}

\IEEEspecialpapernotice{\footnotesize(ICPPW 2016, \copyright IEEE)}

\maketitle

\begin{abstract}
We describe an approach that uses combinatorial optimization and machine learning to share the work between the host and device of heterogeneous computing systems such that the overall application execution time is minimized. We propose to use combinatorial optimization to search for the optimal system configuration in the given parameter space (such as, the number of threads, thread affinity, work distribution for the host and device). For each system configuration that is suggested by combinatorial optimization, we use machine learning for evaluation of the system performance. We evaluate our approach experimentally using a heterogeneous platform that comprises two 12-core Intel Xeon E5 CPUs and an Intel Xeon Phi 7120P co-processor with 61 cores. Using our approach we are able to find a near-optimal system configuration by performing only about 5\% of all possible experiments.
\end{abstract}

\section{Introduction}
\label{sec:introduction}

Heterogeneous computing systems that consist of CPUs and accelerators such as Nvidia GPU \cite{nvidia_tesla} or Intel Xeon Phi \cite{chrysos2014intel} are becoming prevalent. Some of the most powerful supercomputers in the TOP500 list (November 2015,~\cite{top500}) are heterogeneous at their node level. For example, a node of Tianhe-2 (no. 1 in TOP500) comprises two Intel IvyBrigde CPUs and three Intel Xeon Phi co-processors; a node of Titan (no. 2 in TOP500) contains one AMD Opteron CPU and one Nvidia Tesla GPU.

Utilizing the computational power of all the available resources (CPUs + accelerators) in heterogeneous systems is essential to achieve good performance. However, due to different performance characteristics of their processing elements, achieving a good workload distribution across multiple devices on heterogeneous systems is non-trivial \cite{benkner11,vettersurvey,sandrieser12}. Furthermore, optimal workload distribution is most likely to change for different applications, input problem sizes and available resources. Determining the optimal system configuration (including the number of threads, thread affinity, workload partitioning ratio for multi-core processors of the host and the accelerating devices) using brute-force may be prohibitively time consuming.

Various approaches for workload distribution have been proposed. For example Augonnet et al. \cite{starpu2011augonnet} propose a task scheduling library to handle the load balancing and the memory transfer. Scogland et al. \cite{coretsar2014scogland} propose an adaptive worksharing library to schedule computational load across devices. Ravi and Agrawal \cite{ravi2011dynamic} propose a dynamic scheduling framework that splits tasks into smaller ones and distributes them across processing elements on heterogeneous systems. Grewe and O'Boyle \cite{grewe2011static} propose a static partitioning approach to distribute OpenCL programs on heterogeneous systems.

However, so far not much research was focused on using meta-heuristics to optimize the workload distribution of data-parallel applications, which considers various parameters such as: the number of threads, the thread affinity, and the workload partitioning ratio for host CPUs and co-processing devices.

In this paper we propose an optimization approach that combines the Combinatorial Optimization and Machine Learning to determine near-optimal system configuration parameters of a heterogeneous system. We use Simulated Annealing as a combinatorial optimization approach to search for the optimal system configuration in the given parameter space, whereas for performance evaluation of the proposed system configurations during space exploration we use the Boosted Decision Tree Regression. The objective function that we aim to minimize is the application's execution time. To evaluate our approach we use a parallel application for DNA Sequence Analysis on a platform that comprises two 12-core Intel Xeon E5 CPUs and an Intel Xeon Phi 7120P co-processor with 61 cores. Using our optimization approach to determine the near-optimal system configuration we achieve a speedup of 1.74$\times$ compared to the case when only the available resources of the host are used, and up to 2.18$\times$ speedup compared to the case when all the resources of the accelerating device are used.

\textbf{Contributions:} The major contributions of this paper are:
\begin{itemize}
	\item A Combinatorial Optimization approach to explore the large system configuration space;
	\item A supervised Machine Learning approach to evaluate the performance of parallel applications;
	\item An approach that combines the combinatorial optimization heuristic with machine learning to determine a near-optimal system configuration, such that the execution time is decreased;
	\item Experimental evaluation of our approach;
	\item Performance comparison of our approach that utilizes both CPUs and accelerators, compared to CPU-only and accelerator-only approaches.
\end{itemize}

The rest of the paper is organized as follows. Section \ref{sec:background} provides background and motivation. Section \ref{sec:framework} describes the design and implementation of our optimization approach. Section \ref{sec:evaluation} presents our evaluation. This paper is compared and contrasted to the state-of-the-art related work in Section \ref{sec:rw}. We provide conclusions and discuss the future work in Section \ref{sec:conclusion}.

\section{Background and Motivation}
\label{sec:background}

In this section we will motivate the need for optimized workload distribution across heterogeneous devices. To illustrate and motivate the problem of workload distribution on heterogeneous platforms and to evaluate the proposed approach, we will measure the execution time of a DNA Sequence Analysis application \cite{mp-cse15,mp-pbio15} in a heterogeneous platform that is accelerated using an Intel Xeon Phi co-processor. Details related to the heterogeneous platform and the application used for experimentation will follow in the next sections.

\subsection{Heterogeneous Computing Platforms with Intel Xeon Phi}
\label{sec:xphi}
\begin{figure}[tb]
	\centering
	\includegraphics[width=\linewidth]{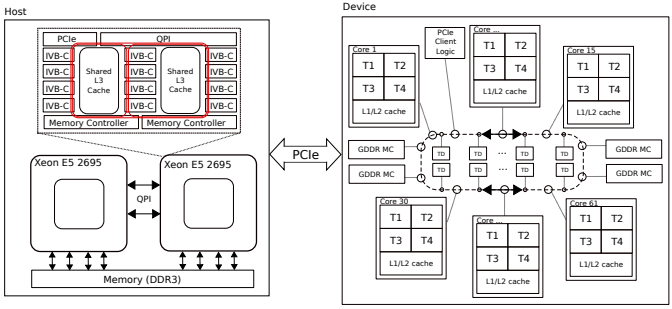}
	\caption{Our target accelerated system comprises a host with two CPUs and an Intel Xeon Phi device.}
	\label{fig:emil-platform}
\end{figure}

A typical heterogeneous platform that is accelerated with the Intel Xeon Phi is diagrammed in Figure \ref{fig:emil-platform}. Such platforms may consist of one or two CPUs on the host (left-hand side of the figure), and one to eight accelerators (right-hand side of the figure). The host CPUs are of type Intel Xeon E5, which consists of 12 cores, each of them supports two hardware threads that amounts to a total of 48 threads. The L3 cache is split in two parts, in total it features a 30MB L3 cache.

The Xeon phi accelerator has 61 cores, where each core supports four hardware threads, in total 244 threads per co-processor \cite{chrysos2014intel}. The Xeon Phi comes with a lightweight Linux Operating System ($\mu$OS) that allows us to either run applications natively or offload them. One of the cores is used by the OS, the remaining 60 cores are used for experimentation. The Xeon Phi has a unified L2 cache memory of 30.5MB. One of the key features of the Intel Xeon Phi is its vector processing units that are essential to fully utilize the co-processor \cite{TianSPGKMCP13}. Through the 512-bit wide SIMD registers it can perform 16 (16 wide $\times$ 32 bit) single-precision  or 8 (8 wide $\times$ 64 bit) double-precision operations per cycle. The performance capabilities of the Intel Xeon Phi have been investigated by different researches within different domains \cite{vp-hpcc15,liu2016parallel,DokulilBBPSB13}.

\subsection{DNA Sequence Analysis}
\label{sec:dna-seq}
For motivation purposes, and later on for evaluation of our approach we have used a high performance data analytic application for DNA Sequence Analysis \cite{mp-cse15,mp-pbio15} that is based on Finite Automata and finds patterns (so called motifs) in large-scale DNA sequences. It allows efficient use of the computational resources of the host and accelerating device. The DNA Sequence Analysis application targets heterogeneous systems that are accelerated with the Intel Xeon Phi co-processor, and is able to exploit both the thread- and SIMD-level parallelism. 

\subsection{Motivational Experiment}
\label{sec:motivation}

We measured the execution time of a DNA Sequence Analysis application \cite{mp-cse15,mp-pbio15} on a simple heterogeneous system that consists of two Intel Xeon CPUs and one Intel Xeon Phi co-processor. In reality, heterogeneous systems may consist of several different types of accelerators with different performance capabilities.
 
We run these experiments with different input sizes and number of CPU threads. To highlight the work-distribution problem we vary the  distribution ratio across host and device. 

\begin{figure}[tb]
	\centering
	\begin{subfigure}[b]{0.45\textwidth}
		\includegraphics[width=\textwidth]{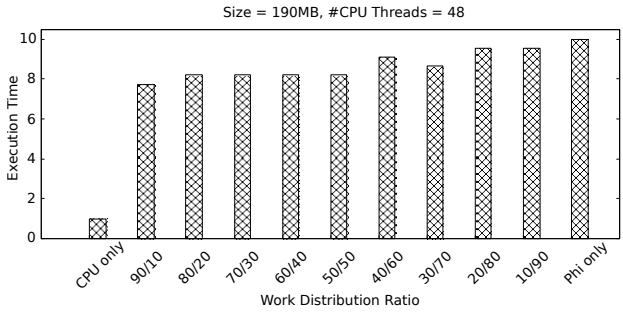}
		\caption{Experiment 1. (Input Size = 190MB, \# CPU Threads = 48)}
		\label{fig:mot-turkey-48}
	\end{subfigure}
	\hfill
	~ 
	\begin{subfigure}[b]{0.45\textwidth}
		\includegraphics[width=\textwidth]{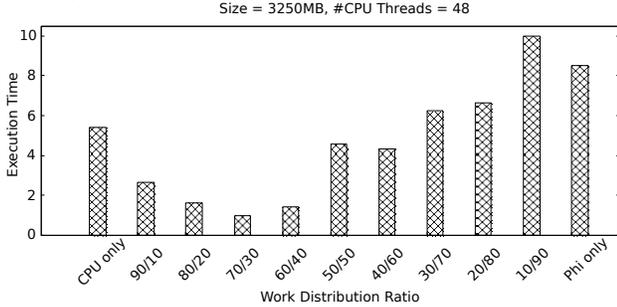}
		\caption{Experiment 2. (Input Size = 3250MB, \# CPU Threads = 48)}
		\label{fig:mot-human-48}
		
	\end{subfigure}
	
	\begin{subfigure}[b]{0.45\textwidth}
		\includegraphics[width=\textwidth]{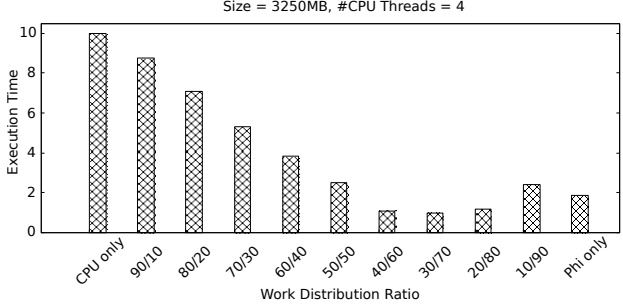}
		\caption{Experiment 3. (Input Size = 3250MB, \# CPU Threads = 4)}
		\label{fig:most-human-4}
	\end{subfigure}

	\caption{DNA Sequence Analysis with different input sizes and number of CPU threads used. The execution time values are normalized in the range of 1-10.}
	\label{fig:mot-case-study}
\end{figure}

Figure \ref{fig:mot-case-study} shows the results of our experiments. The x-axis indicates the work distribution ratio, for instance $60/40$ means that 60\% of the work is mapped to the host CPUs and the remaining 40\% is mapped to the co-processor. The y-axis indicates the execution time, note that the values are normalized in a range from 1-10.  
In the first experiment, depicted in Fig. \ref{fig:mot-turkey-48}, we may observe that the lowest execution time is achieved when running on the CPU only. That is due to the relatively small input size used, where any work distribution makes the execution time be biased by the represented overhead. In the second experiment, shown in Fig. \ref{fig:mot-human-48}, we used a larger input size, therefore running on the 48 threads of the CPU or on the co-processor only is not the most effective mapping. We may observe that a work distribution of $70/30$ or $60/40$ is much faster. Figure \ref{fig:most-human-4} shows the results when using the same input size but the number of CPU threads is reduced to 4. We may observe that the optimal work distribution is when we assign 70\% of the work to the co-processor. Please note that in these experiments we consider only 11 possible workload partition ratios ($0, 10, 20, ..., 100$). In real-world problems this ratio can be any number in the interval 0-100.

From the above experiments we may see that the optimal workload distribution depends on the input size and the available resources. If we consider more features (example, thread affinity, number of threads per core) or multiple accelerators with different performance characteristics, the number of all possible system configurations increases dramatically. Determining the optimal system configuration using brute-force may be prohibitively time expensive. The number of all possible system configurations is a product of parameter value ranges, 

\begin{equation} \label{eq:1}
\prod_{i=1}^{n}	R_{c_i} = R_{c_1} \times R_{c_2} \times .. \times R_{c_n}
\end{equation}

where $C=\{c_1, c_2, ..., c_n\}$ is a set of parameters and each $c_i$ has a value range $R_{c_i}$.

In the next section we are going to propose an intelligent work distribution approach that is able to determine an optimal system configuration using combinatorial optimization and machine learning.

\section{Design and Implementation}
\label{sec:framework}

One of the most compelling features of the Intel Xeon Phi co-processor is the double advantage of transforming-and-tuning, which means that tuning an application on the Intel Xeon Phi for scaling (more cores and threads), vectorization and memory usage, stands to benefit an application when running on the Intel Xeon processors. Therefore, with not much programming investment application tailored for many-core Intel Xeon Phi co-processors can benefit when running on multi-core Intel Xeon CPUs, and vice-versa. To distribute the workload across the heterogeneous devices we use the offload programming model. We overlap the parts offloaded to the co-processor with the ones that are running on the host CPUs, which mitigates the idle time for both CPUs and accelerators.

We target applications with \enquote{divisible} workload, which means that the workloads division can be adjusted arbitrarily. However, as seen in Section \ref{sec:motivation}, in heterogeneous systems that have processing units of different speed, finding an optimal partitioning ratio for a given workload is non-trivial.

In this section, we describe our approach for determining the optimal system configuration parameters (including number of threads, thread affinities, workload fraction) of a heterogeneous systems. The goal of our approach is to propose a near-optimal system configuration such that the overall execution time is minimized. The system parameters and their possible values are listed in Table \ref{table:sys-parameters}. 

\begin{table}[t]
	\centering
	\scriptsize
	\caption{The set of considered parameters and their values for our target system.}
	\label{table:sys-parameters}
	\begin{tabular}{@{}lll@{}}
		\toprule
		\multicolumn{1}{c}{} 	& Host								& Device 									\\ \midrule
		Threads              	& \{2, 4, 6, 12, 24, 36, 48\}      	& \{2, 4, 8, 16, 30, 60, 120, 180, 240\}	\\
		Affinity             	& \{none, scatter, compact\}       	& \{balanced, scatter, compact\}			\\
		Workload Fraction   & \{1..100\}                        & \{100 - Host Workload Fraction\}			\\ \bottomrule
	\end{tabular}
\end{table}

To determine the optimal system configuration in a large parameter space one could try to naively enumerate over all possible parameter values, a technique we refer to as \emph{enumeration} (also known as \emph{brute-force}). The use of enumeration for design-space exploration in a real-world context may be prohibitively time consuming \cite{Bosschere2001,PllanaBMNX08,pbb08,fahringer04}. Therefore, we propose to use Simulated Annealing as a combinatorial optimization method to search for an optimal system configuration in a given parameter space. We may use \emph{measurements} or \emph{model-based prediction} for evaluation of the system performance for each system configuration. In comparison to the measurement based evaluation, the prediction-based is much faster but less accurate. Furthermore it requires training of the prediction model.
In this paper, we consider using various optimization approaches: 

\begin{description}
	\item[a) Enumeration and Measurements (EM)] -
	Certainly determines the optimal system configuration, however it involves a very large number of performance experiments. The expected optimization effort is very high. Since EM has no performance prediction capabilities, for each program input the whole optimization process needs to be repeated.
	
	\item[b) Enumeration and Machine Learning (EML)] - uses machine learning to infer about the system performance. Since it has to examine all of the possible system configurations, the effort needed for parameter space exploration is still high. 

	\item[c) Simulated Annealing and Measurements (SAM)] - uses Simulated Annealing to guide the parameter space exploration and measurements for performance evaluation of the proposed system configurations. This method significantly reduces the effort for parameter space exploration.
	\item[d) Simulated Annealing and Machine Learning (SAML)] - Compared to SAM, SAML provides the possibility to predict the system performance for new unseen system configurations, because it uses machine learning for performance evaluation.
\end{description}

The properties of each of the proposed approaches are listed on Table \ref{table:opt-method-properties}. 
In what follows in this section we describe our approach for parameter space exploration using Simulated Annealing and our approach for performance prediction using Machine Learning.

\begin{threeparttable}[t]
	\scriptsize
	\centering
	\caption{Properties of optimization methods. }
	\label{table:opt-method-properties}
	\begin{tabular}{@{}lllllc@{}}
		\toprule
		Method 	& \specialcell{Space\\Exploration}   & \specialcell{Sys. Conf.\\Evaluation} & Effort		& Accuracy		& Prediction \\ \midrule
		EM      & Enumeration         				 & Measurements			   				& high			& optimal 		& no		 \\
		EML     & Enumeration         				 & \specialcell{Machine\\Learning}		& high     		& near-optimal 	& yes		 \\
		SAM     & \specialcell{Simulated\\Annealing} & Measurements          				& medium   		& near-optimal 	& no		 \\
		SAML    & \specialcell{Simulated\\Annealing} & \specialcell{Machine\\Learning}      & medium 		& near-optimal 	& yes		 \\ \bottomrule
	\end{tabular}
\end{threeparttable}

\subsection{Using Simulated Annealing for Parameter Space Exploration}
\label{sec:SA}

Press et al. \cite{NumRecipes2007} describe several heuristics for solving optimization problems, including: Genetic Algorithms, Ant Colony Optimization, Simulated Annealing, Local Search, Tabu Search. Factors such as the type of the optimization problem and search space, the computational time, and demanded solution quality need to be considered when choosing the most convenient heuristic for a specific problem \cite{ACO4PS, braun2001comparison}. We have decided to use Simulated Annealing because of its ability to cope with very large discrete configuration space, and the ability to avoid getting stuck at local minimums, which makes it much better on average at finding an approximate global minimum on a large space. 

The name and inspiration comes from the process of annealing in metallurgy, a technique that includes heating and controlled cooling of materials. At high temperatures particles of the material have more freedom of movement, and as the temperature decreases the movement of particles is restricted as well. When the material is cooled slowly, the particles are ordered in the form of a crystal that represents its minimal energy.

In the same way, in Simulated Annealing there is a temperature variable $T$ that controls the cooling process. One of the fundamental properties of the Simulated Annealing meta-heuristic is its ability to accept worse solutions at a higher temperature, therefore there is a corresponding chance to get out of local minimum, which enables a more extensive search for the global optimal solution. The lower the temperature, less likely it accepts new solutions \cite{NumRecipes2007}. 

The method of Simulated Annealing is a suitable technique for optimization of large scale problems, especially the ones where the global optimum is hidden among many local optima. Examples like the traveling salesman problem (TSP) or designing complex integrated circuits are just some of many problems that can be solved using the Simulated Annealing. The space over which the objective function is defined is discrete and very large (factorial) configuration space, for example, in the TSP the set of possible orders of cities.

In the context of the load balancing problem in heterogeneous systems, we define the configuration space as follows:

\begin{itemize}
	\item workload fraction is a discrete value from 0-100, which indicates the percentage of the workload that needs to be executed in a specific device. For instance in a heterogeneous system with one CPU and one accelerator, if 40\% of the workload is mapped to the host CPU, the remaining 60\% is assigned to the accelerator(s);
	\item number of threads for the host CPU and the accelerator(s);
	\item the thread allocation strategy for the host CPU and the accelerator(s);
\end{itemize}

The objective function $E$ (analog of energy) of our approach is to minimize the total execution time of an application, which basically is determined by the maximum of the $T_{host}$ and $T_{device}$:

\begin{equation}
E = max(T_{host}, T_{device})
\end{equation}

\begin{figure}[tb]
	\centering
	\includegraphics[width=.7\linewidth]{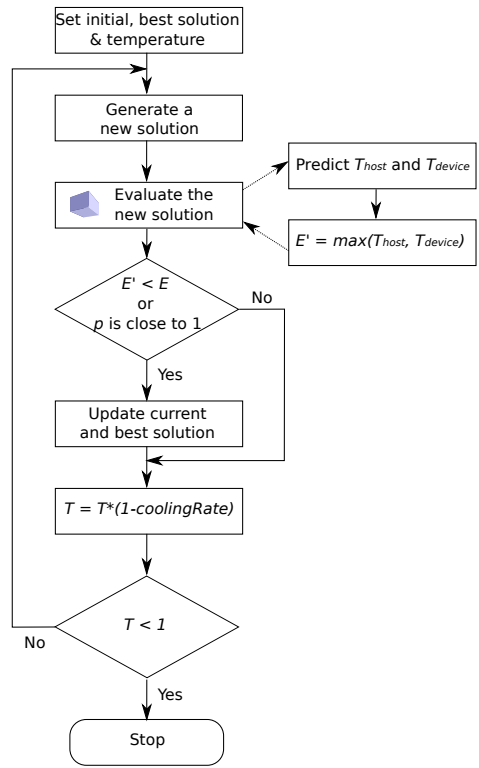}
	\caption{The Structure of the Simulated Annealing Algorithm.}
	\label{fig:sa-algorithm}
\end{figure}

An overview of the Simulated Annealing algorithm is depicted in Figure \ref{fig:sa-algorithm}. The algorithm start by setting an initial temperature and creating a random initial solution. Then we begin looping until the annealing process has sufficiently cooled. We define the annealing schedule as follows:

\begin{equation}
T =  T * (1 - coolingRate);
\end{equation}

where $coolingRate$ determines the cooling rate.

The temperature variable plays a decisive role in the acceptance probability function. When a new solution is proposed, we first check if its energy $E'$ is lower than the energy of the current solution $E$. If it is, we accept it unconditionally, otherwise we consider how much worse is the time of the proposed solution compared to the current one, and what is the temperature of the system. If the temperature is high, the system is more likely to accept solutions that are worse than the current one. The acceptance probability function $p$ is determined as follows:

\begin{equation}
p = exp((E - E') / T)
\end{equation}

where $E'$ determines the energy of the newly generated solution. This function allows the system to get out of local optima, and find a new better one. 

\subsection{Using Machine Learning for Performance Evaluation}
\label{sec:ml}

The evaluation of the newly generated solutions by the Simulated Annealing can be done using measurements of actual program execution, or using machine learning approaches to predict the execution time of an application on the host $T_{host}$ and accelerator $T_{device}$. In our approach we use the predicted execution time to determine the near-optimal system configuration. The aim is to balance the workload between the host and device(s) such that the total execution time is reduced.

During the development of our performance prediction model we have considered various supervised machine learning approaches, including Linear Regression, Poisson Regression, and the Boosted Decision Tree Regression. In our performance prediction experiments, we achieved more accurate prediction results with the Boosted Decision Tree Regression. The Boosted Decision Tree Regression is a supervised machine learning algorithm that uses boosting to generate a group of regression trees and determine the optimal tree based on a loss function. 

\begin{figure}[tb]
	\centering
	\includegraphics[width=\linewidth]{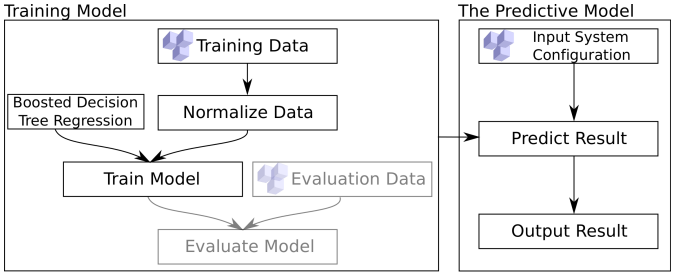}
	\caption{The Predictive Model using Boosted Decision Tree Regression}
	\label{fig:ml-algorithm}
\end{figure}

The execution time for most of the applications is mainly influenced by the input size, the available computing resources, and the thread allocation strategies. Therefore, we use these features to train and evaluate our prediction model. 

We have generated training data for training our performance prediction model by executing the application used during evaluation of our approach with different number of threads, thread affinities and input sizes. The main features including their possible values used to train and evaluate our prediction model are listed in Table \ref{table:sys-parameters}.

We generated data by running our experiments on two different environments (host and device). On the host we used $2, 4, 6, 12, 24, 36$ and $48$ threads. We varied the thread affinities between \emph{none, scatter,} and \emph{compact}. On the accelerator we used $2, 4, 8, 16, 30, 60, 120, 180$ and $240$ threads, whereas we varied the thread affinity strategies between \emph{balanced, scatter,} and \emph{compact}. We trained our model with different input fractions, varying from 0-100, which represents the percentage fraction of the input that needs to be examined in a specific device. In total the data of about 7200 experiments were used to train and evaluate the performance prediction model using the Boosted Decision Tree Regression. Half of the experiments were used to train the prediction model, and the other half were used for evaluation.

Figure \ref{fig:ml-algorithm} illustrates the process of training and predicting an unseen system configuration. The left hand side of the figure shows the training model, which basically takes as input a structured data set, and trains a model using the Boosted Decision Tree Regression algorithm. The gray colored boxes are used for evaluation of our approach. The right-hand side of the figure shows the Predictive Model, which takes the proposed system configurations as input, uses the trained model and predicts the execution time.

\section{Evaluation}
\label{sec:evaluation}

In this section we evaluate experimentally our proposed combinatorial optimization approach for workload distribution on heterogeneous platforms. We describe the following:

\begin{itemize}
	\item the experimentation environment
	\item evaluation of our prediction model
	\item comparison of the SAML and EM
	\item achieved performance improvement
\end{itemize}

\subsection{Experimentation Environment}
\label{sec:exp-env}

In this section we describe the experimentation environment used for the evaluation of our approach for workload sharing on heterogeneous platforms. We describe the system configuration, the application used for testing, its input dataset, and the parameters that define the system configuration.

In Section \ref{sec:xphi} we described the architecture of the heterogeneous platform used for performance evaluation of our approach. The major features of our system are listed in Table \ref{table:emil}. In Section \ref{sec:dna-seq} we talked about the application used for evaluation of our approach, that is a DNA Sequence Analysis application. We used the code generated by our PaREM tool \cite{mp-cse14} as a basis for our DNA Sequence Analysis application.

The DNA sequence is basically a long string of characters. Each character indicates one of the nucleotide bases Adenine (A), Cytosine (C), Guanine (G), and Thymine (T). The size of the DNA sequences of various organisms is typically of several gigabytes. For experimentation, we used real-world DNA sequences of human (3.17GB), mouse (2.77GB), cat (2.43GB) and dog (2.38GB). These DNA sequences are extracted from the GenBank sequence database of the National Center for Biological Information \cite{GenBank}.

The parameters that define the system configuration for our combinatorial optimization approach are shown in Table \ref{table:sys-parameters}. All the parameters are discrete. The considered values for the number of threads for host are $\{2, 6, 12, 24, 36, 48\}$, whereas for device are $\{2, 4, 8, 16, 30, 60, 120, 180, 240\}$. The thread affinity can vary between $\{none, compact, scatter\}$ for the host, and $\{balanced, compact, scatter\}$ for the device. The DNA Sequence Fraction parameter can have any number in the range $\{0,..,100\}$, such that if $60\%$ of the DNA sequence is assigned for processing to the host, the remaining $100 - 60 = 40\%$ is assigned to the device.

\begin{table}[!t]
	\scriptsize
	\caption{\emph{Emil}: hardware architecture}
	\label{table:emil}
	\centering
	\begin{tabular}{lll}
		\toprule
		Specification 				   & Intel Xeon 		  & Intel Xeon Phi \\ \midrule
		Type 						   & E5-2695v2 			  & 7120P 	  	   \\ 
		Core frequency                 & 2.4 -- 3.2 GHz       & 1.238 -- 1.333 GHz	   \\ 
		\# of Cores                    & 12                   & 61        	   \\ 
		\# of Threads                  & 24                   & 244       	   \\ 
		Cache                          & 30 MB                & 30.5 MB   	   \\ 
		Max Mem. Bandwidth	           & 59.7 GB/s            & 352 GB/s  	   \\ 
		Memory                         & 8x16 GB              & 16 GB     	   \\ \bottomrule
	\end{tabular}
\end{table}

\subsection{Evaluation of our Performance Prediction Model}
\label{sec:eval-ml}

We have trained our performance prediction model for different input sizes. A total of 7200 experiments (2880 on host and 4320 on the device) were performed. We employed a standard validation methodology by using half of the experiments for training and the other half for evaluation. The predicted values are then compared to the measured values to calculate the prediction accuracy. We use the absolute error and the percent error to express the prediction accuracy, 

\begin{equation}
absolute\_error = |T_{measured} - T_{predicted}|
\end{equation}

\begin{equation}
percent\_error = 100 \cdot absolute\_error / T_{measured}
\end{equation}

\textbf{Result 1:} \textit{The execution times evaluated by our performance prediction model match well the execution time evaluated with measurements. }

\begin{figure}[!h]
	\centering
	\includegraphics[width=\linewidth]{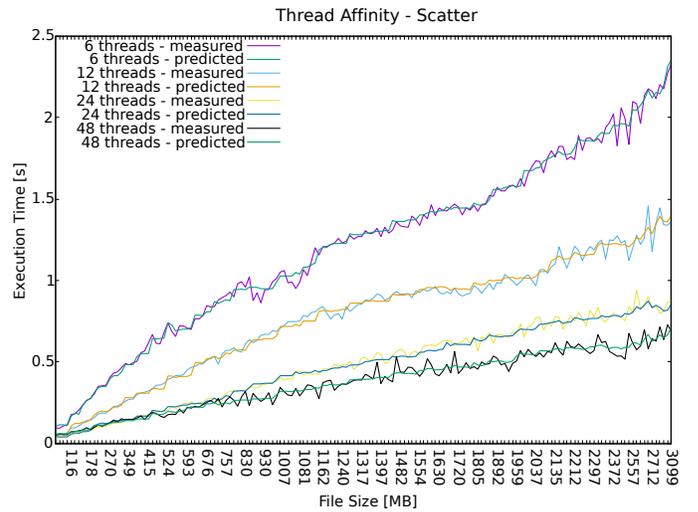}
	\caption{Performance prediction accuracy for the host. A total of 2880 experiments with DNA sequences of human, mouse, cat and dog were needed. Half of the experiments are used to train the model, and the other half to evaluate it.}
	\label{fig:host-scatter}
\end{figure}

Figure \ref{fig:host-scatter} shows the measured and predicted execution time of DNA sequence analysis on the host CPUs. We perform the experiments for various number of threads, thread affinities, and fractions  of the selected DNA sequences. The fractions include $2.5 - 100$ percent of the DNA sequence size. We may observe that predicted values match well the measured values execution times for most configurations. We observe similar behavior for \emph{none} and \emph{compact} thread affinities, but we elide these figures for space and simplicity.

\begin{figure}[!h]
	\centering
	\includegraphics[width=\linewidth]{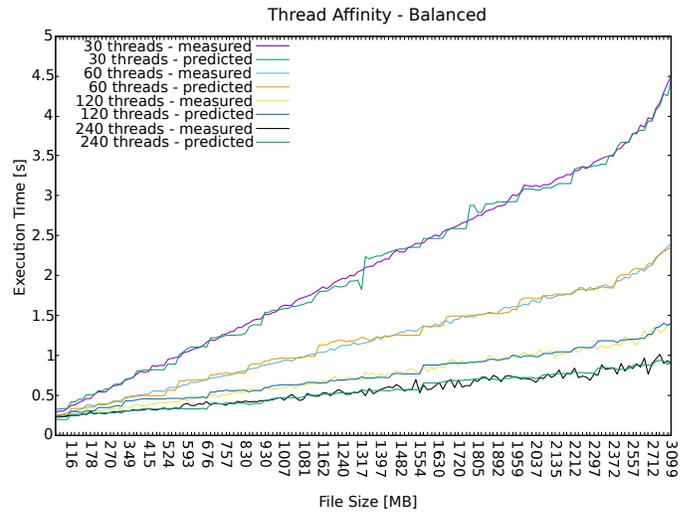}
	\caption{Performance prediction accuracy for the device. A total of 4320 experiments with DNA sequences of human, mouse, cat and dog were needed. Half of the experiments are used to train the model, and the other half to evaluate it.}
	\label{fig:device-balanced}
\end{figure}

Figure \ref{fig:device-balanced} depicts the measurement and prediction results of the execution time on the Intel Xeon Phi device for different number of threads and fractions of the selected DNA sequences. For most of the test cases the predicted execution time values match well the measured values. We have observed similar behavior when using 2, 4, 8, and 16 threads and varying the thread affinity to \emph{scatter} and \emph{compact}, but we elide their results for space and simplicity. 

\begin{table}[h]
	\centering
	\caption{Performance prediction accuracy expressed via the absolute error [s] and percent error [\%] for the host}
	\label{table:error-ml-host}
	\begin{tabular}{@{}llllllll@{}}
		\toprule
		Threads          & 2     & 6     & 12    & 24    & 36    & 48    & avg \\ \midrule
		absolute {[}s{]} & 0.032 & 0.032 & 0.027 & 0.026 & 0.023 & 0.023 & 0.027   \\
		percent {[}\%{]} & 1.756 & 4.102 & 5.678 & 7.141 & 6.555 & 6.201 & 5.239   \\ \bottomrule
	\end{tabular}
\end{table}

\begin{table}[]
	\centering
	\caption{Performance prediction accuracy expressed via the absolute error [s] and percent error [\%] for the device}
	\label{table:error-ml-device}
	\begin{tabular}{@{}p{1.32cm} p{0.3cm} p{0.3cm} p{0.3cm} p{0.3cm} p{0.3cm} p{0.3cm} p{0.3cm} p{0.3cm} p{0.3cm} p{0.7cm}@{}}
		\toprule
		Threads          & 2     & 4     & 8     & 16    & 30    & 60    & 120   & 180   & 240   & avg \\ \midrule
		absolute {[}s{]} & 0.16 & 0.16 & 0.11 & 0.06 & 0.05 & 0.04 & 0.03 & 0.03 & 0.03 & 0.074   \\
		percent {[}\%{]} & 1.21 & 1.98 & 2.68 & 2.56 & 2.92 & 3.54 & 4.38 & 4.22 & 4.68 & 3.132   \\ \bottomrule
	\end{tabular}
\end{table}

\textbf{Result 2:} \textit{The performance prediction model is able to accurately predict the execution time for unseen system configurations. The absolute and percent error are very low.}

\begin{figure}[tb]
	\centering
	\includegraphics[width=\linewidth]{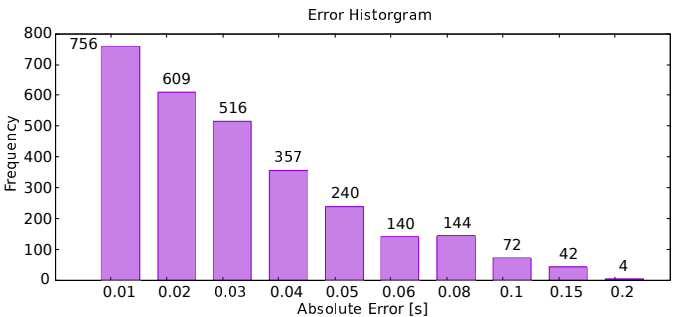}
	\caption{Error histogram for execution time predictions on the host}
	\label{fig:host-histogram}
\end{figure}

Figure \ref{fig:host-histogram} depicts a histogram of the frequency of performance prediction absolute error for the experiments running on the host CPUs. It shows that most of the absolute error values are low. For instance, $756$ predictions have an absolute error less than $0.01$ seconds, $609$ predictions have an absolute error in the range $0.01 - 0.02$ seconds, and the rest of the predictions have an absolute error in the range of $0.02 - 0.2$.

\begin{figure}[tb]
	\centering
	\includegraphics[width=\linewidth]{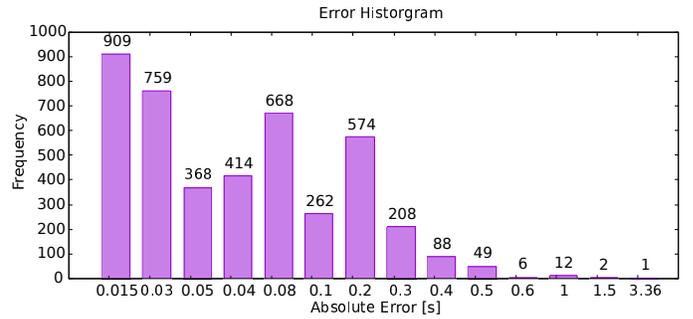}
	\caption{Error histogram for execution time predictions on the device}
	\label{fig:device-histogram}
\end{figure}

Figure \ref{fig:device-histogram} depicts a histogram of the frequency of performance prediction absolute errors for the experiments running on the co-processor. Most of the predictions have an absolute error less than $0.3$ seconds. The error differences between the host and device error histograms is due to the larger span of execution times (0.9 - 42 seconds) on the device compared to host (0.74 - 5.5 seconds). However, that does not necessarily mean that the prediction model for the device is less accurate than the one on for the host (see the percent errors in Table \ref{table:error-ml-host} and \ref{table:error-ml-device}).

The average percent and absolute error that considers all the tested system configurations for different number of threads on the host is shown in Table \ref{table:error-ml-host}. Table \ref{table:error-ml-device} shows the average percent and absolute error for the experiments running on the co-processor. The average percent error for the experiments on the host is 5.239\%, whereas the average percent error on the device is 3.132 \%. The average absolute error on the host is 0.027 seconds, and 0.074 on the device.

In the following section, we will show that the average prediction error of 5.239\% and 3.132\% enables us to satisfactory infer about the execution time during the evaluation of a given system configuration.

\subsection{Comparison of SAML with EM}
\label{sec:comparison}

The enumeration approach finds the system parameter values that result with the best performance by trying out all of the possible parameter values of the system under study. While this approach determines certainly the best system configuration, for the large search space of real-world problems enumeration may be prohibitively expensive. For the experiments used in this paper, despite the fact that we tested only what we considered reasonable parameter values (listed on Table \ref{table:sys-parameters} in Section \ref{sec:framework}), 19926 experiments were required when we used enumeration. Our heuristic-guided approach SAML that is based on Simulated Annealing and Machine Learning leads to comparatively good performance results, which requires only a relatively small set of experiments to be performed. 

For performance comparison, we use the absolute difference and percent difference, which are determined using the following equations:

\begin{equation}
absolute\_difference = |T_{EM} - T_{SAML}|
\end{equation}

\begin{equation}
percent\_difference = 100 \cdot absolute\_difference / T_{EM}
\end{equation}

where $T_{EM}$ indicates the best execution time determined using EM, and $T_{SAML}$ indicates the execution time of our algorithm with a system configuration suggested by the SAML approach. 

\textbf{Result 3} \textit{Using SAML we can determine a near-optimal system configuration by evaluating only about 5\% of the total required experiments by EM}

\begin{figure*}[bt]
	\centering
	\begin{subfigure}[b]{0.47\textwidth}
		\includegraphics[width=\textwidth]{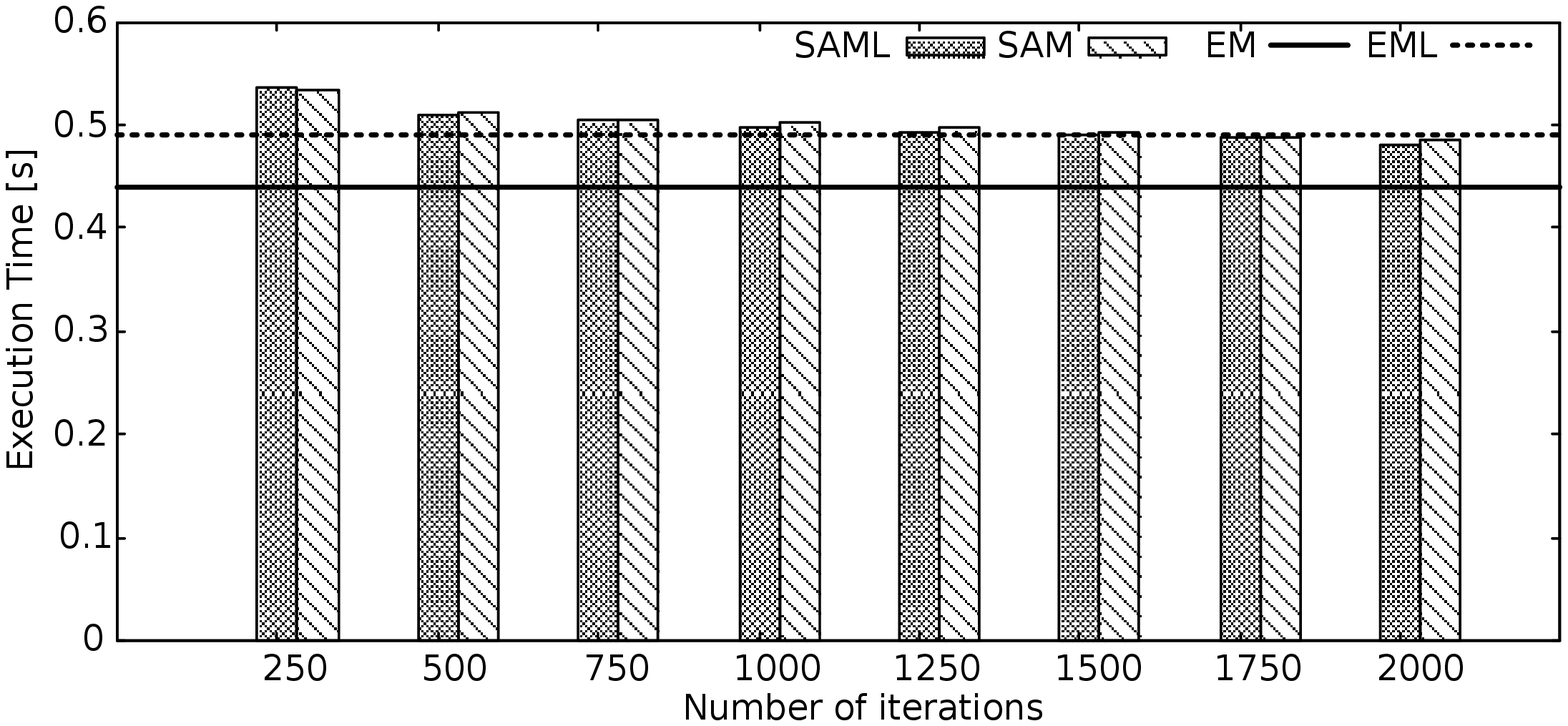}
		\caption{the sequence of human}
		\label{fig:samlhuman}
	\end{subfigure}
	\hfill
	~ 
	\begin{subfigure}[b]{0.47\textwidth}
		\includegraphics[width=\textwidth]{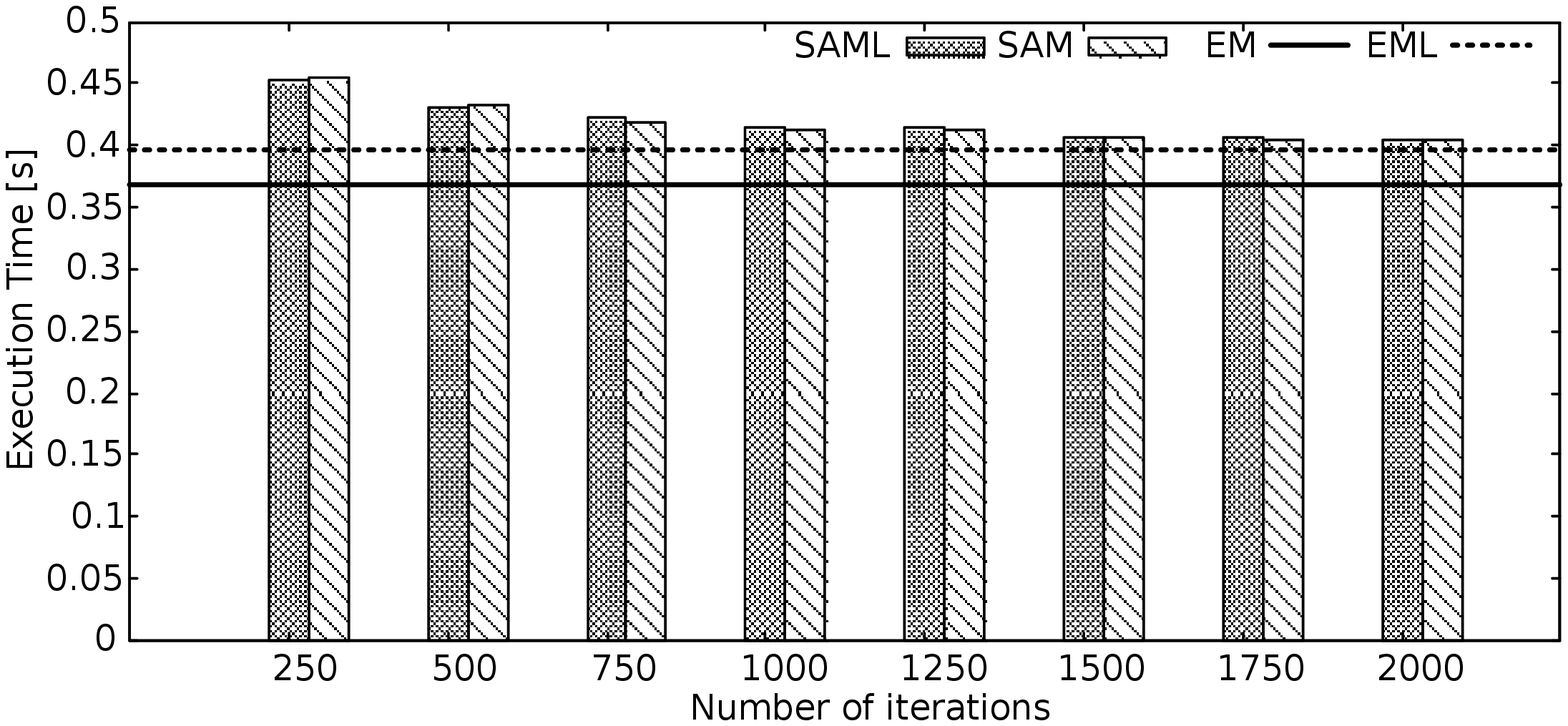}
		\caption{the sequence of mouse}
		\label{fig:saml_mouse}
	\end{subfigure}
	\begin{subfigure}[b]{0.47\textwidth}
		\includegraphics[width=\textwidth]{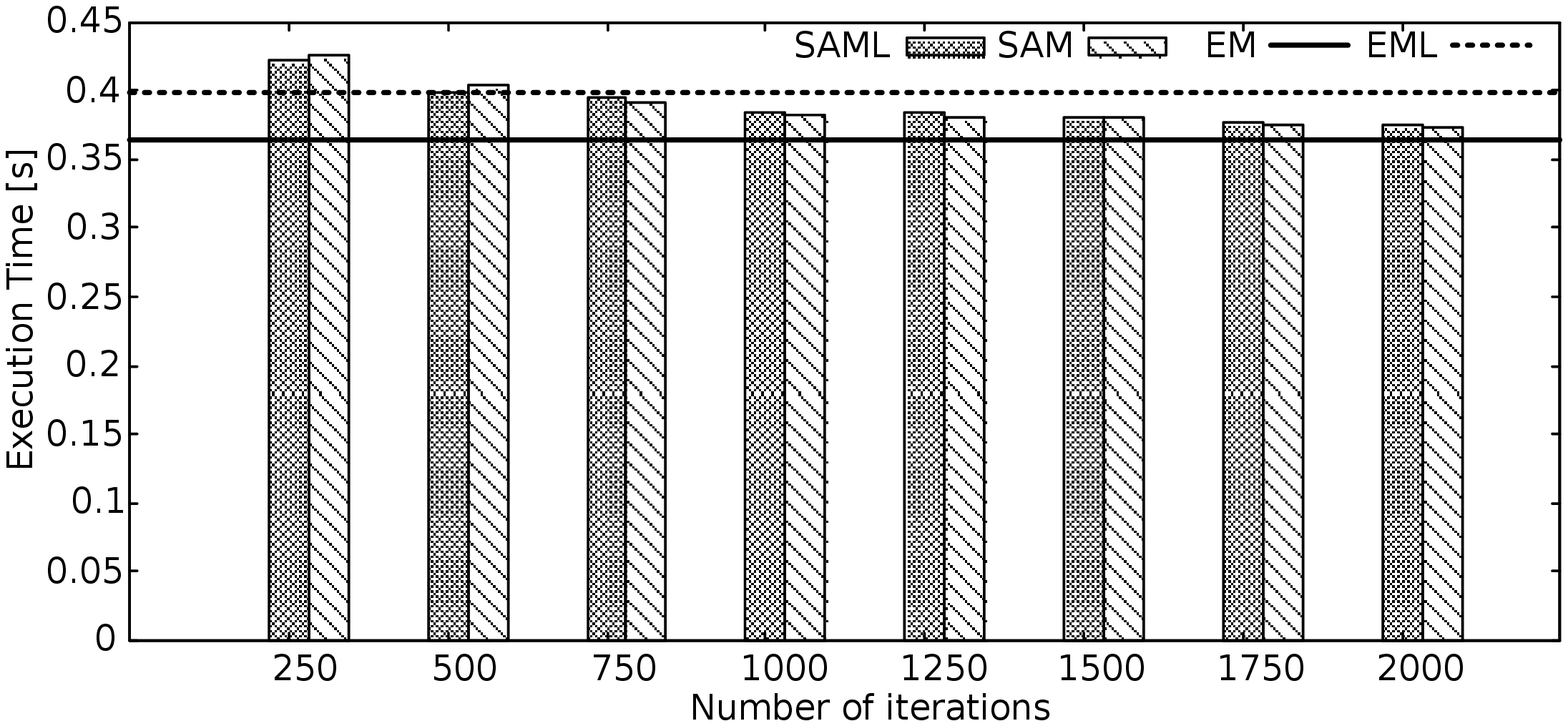}
		\caption{the sequence of cat}
		\label{fig:saml_cat}
	\end{subfigure}
	\hfill
	~ 
	\begin{subfigure}[b]{0.47\textwidth}
		\includegraphics[width=\textwidth]{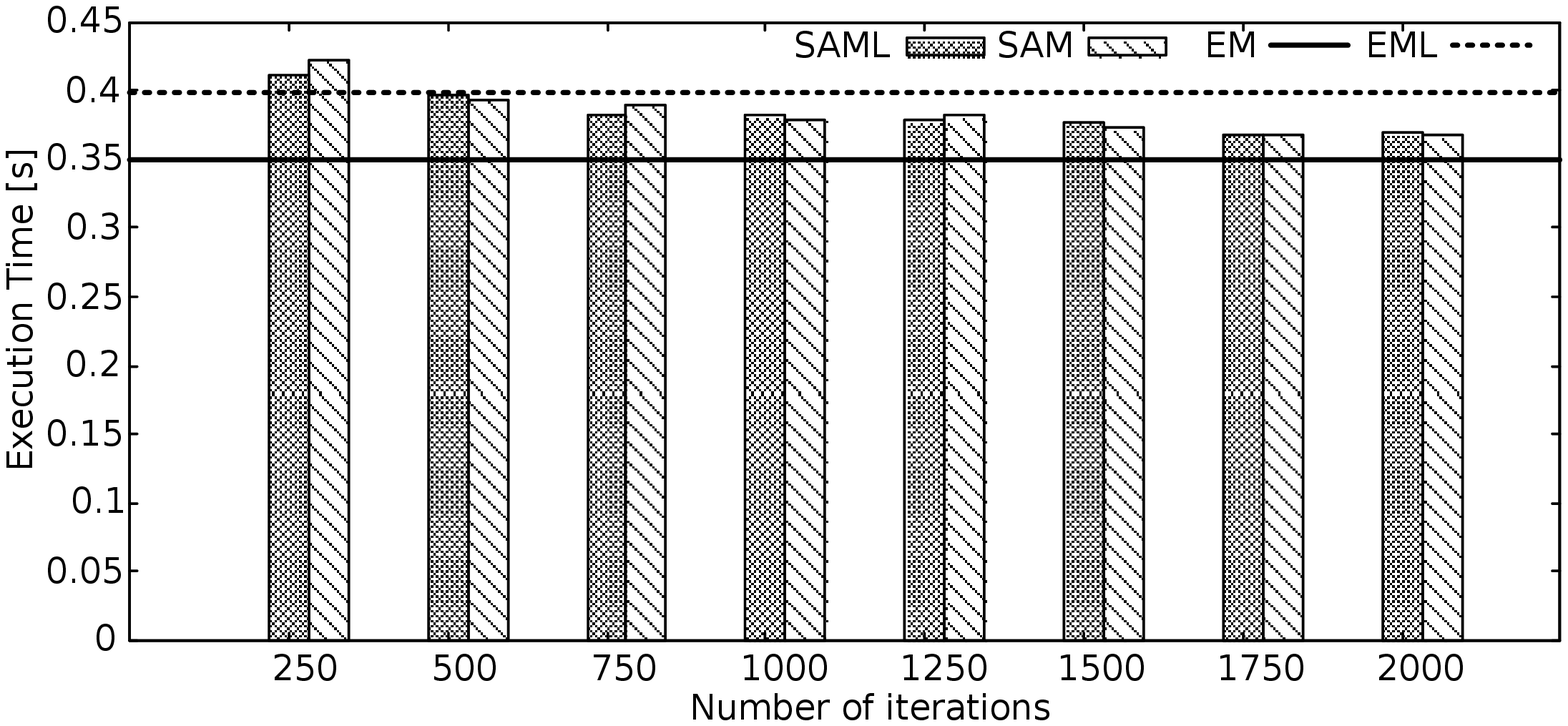}
		\caption{the sequence of dog}
		\label{fig:saml_dog}
	\end{subfigure}	
	\caption{Performance comparison between the best system configuration determined by the Enumeration and Measurements (EM) and the near to optimal one determined by the Simulated Annealing and Measurements (SAM) and Simulated Annealing and Machine Learning (SAML).}
	\label{fig:saml-enum}
\end{figure*}

Figure \ref{fig:saml-enum} depicts the execution time of the selected application when running using the system configuration suggested by the simulated annealing. The solid horizontal line indicates the execution time of the system configuration determined by EM, which is considered as the optimal solution. The dashed horizontal line indicates the execution time of the optimal solution determined using EML. 

Simulated Annealing suggests at each \emph{iteration} parameter values for the system configuration. We can adjust the number of iterations required by Simulated Annealing by changing the initial temperature, or adjusting the cooling function. We may observe that after 1000 iterations (that is only about 5\% of the total possible configurations) our approach is able to determine a system configuration that results with a performance that is close to the performance of the system configuration determined with 19926 experiments when using EM. Please note that Simulated Annealing is a global optimization approach, and to avoid ending at a local optima during the search sometimes it accepts a worse system configuration that results with a higher execution time compared to the previous one.

The EML and SAML use the predicted execution times to evaluate the proposed system configurations during the search space, however for fair comparison we use the measured values. That explains the results depicted on Figure \ref{fig:saml_cat} and \ref{fig:saml_dog}, where the execution time for EML is worse than the SAM and SAML for 750 or more iterations. In these cases, based on the predicted values the optimal execution time would be the ones indicated by the dashed lines, however they might be the cases with lowest prediction accuracy. 

\textbf{Result 4} \textit{The system configurations determined using the SAML approach have low absolute and percent differences compared to the optimal solution determined by EM}

Table \ref{table:percent-error-saml} shows the percent difference of the SAML approach compared to the EM. We may observe that for 250 iterations the average percent difference is very high (19.685\%), but by increasing the number of iterations to 500, 750 and 1000, the percent difference decreases significantly, into 14.067\%, 11.846\% and 10.129\% respectively. Further increase of the number of iterations (1250, 1500, 1750 and 2000) results with a modest decrease of the percent difference (9.557, 8.599, 7.601, 6.849). However, since SAML is based on performance predictions, once the model is trained one can easily increase the number of iterations even more in order to achieve a higher accuracy.

With respect to the absolute difference shown in Table \ref{table:absolute-error-saml}, the determined system configurations using SAML with 250 iterations is only 0.075 seconds slower than the EM approach. Increasing the number of iterations into 500, 750 and 1000, decreases the absolute difference between the execution time into 0.054, 0.046 and 0.039 seconds respectively. Doubling the number of iterations required by SAML, we may achieve even closer absolute difference between EM and SAML, only 0.026 seconds. 

\begin{table}[]
	\centering
	\scriptsize
	\caption{Percent difference [\%]. The performance of system configuration suggested by SAML after 250, 500, 750, 1000, 1250, 1500, 1750, 2000 iterations is compared with the best one determined by EM.}
	\label{table:percent-error-saml}
	\begin{tabular}{@{}l| llllllll@{}}
		\toprule
		& \multicolumn{7}{c}{System Configuration} \\
		DNA											& 250   & 500   & 750    & 1000   & 1250   & 1500   & 1750   & 2000	 \\ \midrule
		human										& 22.15 & 16.17 & 14.59 & 13.22   & 12.11  & 11.44  & 11.06  & 9.324   \\
		mouse		  								& 22.80 & 16.84 & 14.47 & 12.25   & 12.28  & 10.50  & 10.35  & 9.488   \\
		cat  										& 15.81 & 9.524 & 8.71  & 5.771   & 5.607  & 4.453  & 3.385  & 2.895   \\
		dog 		  								& 17.98 & 13.74 & 9.61  & 9.269   & 8.233  & 7.998  & 5.613  & 5.691   \\ \midrule
		\textbf{\specialcell{avgerage\\difference}} & 19.68 & 14.07 & 11.85 & 10.13   & 9.557  & 8.599  & 7.601  & 6.849   \\ \bottomrule
	\end{tabular}
\end{table}

\begin{table}[]
	\centering
	\scriptsize
	\caption{Absolute difference [s]. The performance of system configuration suggested by SAML after 250, 500, 750, 1000, 1250, 1500, 1750, 2000 iterations is compared with the best one determined by EM.}
	\label{table:absolute-error-saml}
	\begin{tabular}{@{}l|llllllll@{}}
		\toprule
		& \multicolumn{7}{c}{System Configuration} \\
		DNA							& 250   & 500   & 750    & 1000   & 1250   & 1500   & 1750   & 2000	    \\ \midrule
		human		 				& 0.097 & 0.071 & 0.064  & 0.058  & 0.053  & 0.050  & 0.049  & 0.041	\\
		mouse 		 				& 0.084 & 0.062 & 0.053  & 0.045  & 0.045  & 0.038  & 0.038  & 0.035	\\
		cat 		    			& 0.057 & 0.035 & 0.032  & 0.021  & 0.020  & 0.016  & 0.012  & 0.010	\\
		dog			    			& 0.063 & 0.048 & 0.034  & 0.032  & 0.029  & 0.028  & 0.019  & 0.019	\\ \midrule
		\textbf{\specialcell{average\\difference}} & 0.075 & 0.054 & 0.046  & 0.039  & 0.037  & 0.029  & 0.029  & 0.026    \\ \bottomrule
	\end{tabular}
\end{table}

\subsection{Performance improvement}
\label{sec:performance-improvement}

In this section we present the performance improvement when all the available resources of the host and device are utilized using the system configuration determined by the SAML approach. Please note that in what follows we present only the speedups achieved when comparing our approach with CPU-only (48 threads) and accelerator-only (244 threads) execution times. Comparing our approach with sequential execution is not relevant for this paper.

\textbf{Result 5} \textit{Our approach is able to determine system configurations that allow the applications to efficiently share its workload among the available resources.}

The results in Table \ref{table:speedup-vs-host} demonstrate the performance improvement achieved when the system configuration determined by the SAML and EM is used for DNA sequence analysis compared to the case when all the available cores on the host are used. We achieve a maximal speedup of 1.74 after 1000 system configurations have been tried with SAML, whereas the maximal speedup that can be achieved using EM is 1.95.

\begin{table}[ht]
	\centering
	\scriptsize
	\caption{Speedup achieved when host and device are used for DNA sequence analysis compared with the host only. We consider system configurations determined by EM and SAML after 250, 500, 750, 1000, 1250, 1500, 1750, 2000 iterations.}
	\label{table:speedup-vs-host}
	\begin{tabular}{@{}l|lllllllll@{}}
		\toprule
		& \multicolumn{8}{c}{System Configuration} \\
		DNA   & 250  & 500  & 750  & 1000 & 1250 & 1500 & 1750 & 2000 & EM\\ \midrule
		human & 1.37 & 1.45 & 1.46 & 1.49 & 1.5  & 1.51 & 1.52 & 1.53 & 1.68 \\
		mouse & 1.6  & 1.66 & 1.7  & \textbf{1.74} & 1.75 & 1.77 & 1.77 & 1.78 & \textbf{1.95} \\
		cat   & 1.5  & 1.58 & 1.62 & 1.66 & 1.68 & 1.7  & 1.7  & 1.7  & 1.76 \\
		dog   & 1.42 & 1.51 & 1.52 & 1.56 & 1.57 & 1.58 & 1.6  & 1.6  & 1.69 \\ \bottomrule
	\end{tabular}
\end{table}

Table \ref{table:speedup-vs-device} shows the performance improvement that is achieved when the system configuration determined by the SAML and EM is used for DNA sequence analysis compared to the case when all the available cores on the device are used. The maximal achieved speedup using EM is 2.36. We achieve a close to maximal speedup (2.18) using only 1000 iterations.

\begin{table}[ht]
	\centering
	\scriptsize
	\caption{Speedup achieved when host and device are used for DNA sequence analysis compared with the device only. We consider system configurations determined by EM and SAML after 250, 500, 750, 1000, 1250, 1500, 1750, 2000 iterations.}
	\label{table:speedup-vs-device}
	\begin{tabular}{@{}l|lllllllll@{}}
		\toprule
		& \multicolumn{8}{c}{System Configuration} \\
		DNA	  & 250  & 500  & 750  & 1000 & 1250 & 1500 & 1750 & 2000 & EM\\ \midrule
		human & 1.64 & 1.74 & 1.76 & 1.79 & 1.81 & 1.81  & 1.83 & 1.84 & 2.02 \\
		mouse & 1.7  & 1.77 & 1.80 & 1.85 & 1.86 & 1.88 & 1.88 & 1.89 & 2.07 \\
		cat   & 1.96 & 2.08 & 2.13 & \textbf{2.18} & 2.21 & 2.24 & 2.23  & 2.24 & 2.31 \\
		dog   & 1.99 & 2.1  & 2.13 & \textbf{2.18} & 2.19 & 2.21 & 2.23  & 2.25  & \textbf{2.36} \\ \bottomrule
	\end{tabular}
\end{table}

\section{Related Work}
\label{sec:rw}

Efficient utilization of the combined computation power of the various computing units in heterogeneous systems requires optimal workload distribution. Recent related work proposed various approaches for workload distribution across different devices in heterogeneous systems.

CoreTsar \cite{coretsar2014scogland} is an adaptive worksharing library for workload scheduling across different devices. It is a directive based library that extends the accelerated OpenMP by introducing a cross-device worksharing directive. Such directives enable the programmer to specify the association between the computation and data. The library evaluates the speed of each device statically, then use these indicators to split the workload across different devices. Similarly Ayguad{\'e} et al. \cite{ayguade2003schedule} investigated the extension of OpenMP to allow workload distribution on future iterations based on the results of first static ones. These approaches tend to minimize the required source code changes.

In comparison, StarPU \cite{starpu2011augonnet} and OmpSs \cite{ompss2011duran} (task block models) require manual workload distribution by the developer, which may include significant structural source code changes. These powerful models for scheduling on heterogeneous systems are queue-based that basically split the workload into smaller tasks and queuing these tasks across the available resources. A similar approach based on priority queues is proposed by Dokulili et al. \cite{DokulilBBPSB13}.

A dynamic scheduling framework that divides tasks into smaller ones is proposed by Ravi and Agrawal \cite{ravi2011dynamic}. These task are distributed across different processing elements in a task-farm way. While making scheduling decisions, architectural trade-offs, computation and communication patterns are considered. Our approach considers only system runtime configuration and the input size that makes it a more general approach, which can be used with different applications and architecture.

Odajima et al. \cite{XcalableMP-dev} combines the pragma-based XcalableMP (XMP) \cite{xcalablemp2010nakao} programming language with StarPU runtime system to utilize resources on each heterogeneous node for work distribution of the loop executions. XMP is used for work distribution and synchronization, whereas StarPU is used for task scheduling.

Qilin \cite{qilin2009luk} is a programming system that is based on a regression model to predict the execution time of kernels. Similarly to our approach, it uses off-line learning that is thereafter used in compile time to predict the execution time for different input size and system configuration. 

Grewe and O'Boyle \cite{grewe2011static} focus on workload distribution of OpenCL programs on heterogeneous systems. Their static based partitioning uses static analysis for code features extraction, which are used to determine the best partitioning across the different devices. Their approach relies on the architectural characteristics of a system.

In comparison to the aforementioned approaches, in addition to using machine learning for evaluation of applications performance, we use combinatorial optimization to determine the near-optimal system configuration.

\section{Summary and Future Work}
\label{sec:conclusion}
In this paper we have proposed a combinatorial optimization approach that uses machine learning to determine the system configuration (that is, the number of threads, thread affinity, and the DNA sequence fraction for the host and device) such that the overall execution time is minimized.

We have observed that searching for the best system configuration using enumeration is time consuming, since it required many experiments. Using Simulated Annealing to suggest at each \emph{iteration} parameter values for the system configuration after 1000 iterations we determined a system configuration that results with a performance that is close to the performance of the system configuration determined with 19926 experiments of enumeration. By running only about 5\% of experiments we were able to find a near-optimal system configuration. 

Furthermore, we have proposed a Machine Learning approach that is able to predict the execution time for a system configuration. We have observed in our experiments that the average percent error of 4.2\% (5.239\% on the host, and 3.132\% on the device) of the performance prediction enables us to satisfactory suggest near to optimal system configurations. Using the near optimal system configuration determined by the Simulated Annealing and Machine Learning we achieved a maximal speedup of $1.74\times$ compared to the case when all the cores of the host are used, and up to $2.18\times$ faster compared to the fastest execution time on the device.

Future work will study adaptive workload-aware approaches.


\end{document}